\documentclass{epl}

\usepackage{amsmath}

\usepackage{txfonts}

\title{Monte-Carlo simulation of string-like colloidal assembly}
\shorttitle{String-like colloidal assembly}

\author{Yuki Norizoe \and Toshihiro Kawakatsu}
\shortauthor{Yuki Norizoe \etal}

\institute{                    
  Department of Physics, Tohoku University - Sendai 980-8578, Japan
}


\begin{document}

\maketitle

\begin{abstract}
We study structural phase transition of polymer-grafted colloidal particles by Monte Carlo simulations on hard spherical particles. The interaction potential, which has a weak repulsive step outside the hard core, was validated with use of the self-consistent field calculations. With this potential, canonical Monte Carlo simulations have been carried out in two and three dimensions using the Metropolis algorithm. At low temperature and high density, we find that the particles start to self-assemble and finally align in strings. By analyzing the cluster size distribution and string length distribution, we construct a phase diagram and find that this string-like assembly is related to the percolation phenomena. The average string length diverges in the region where the melting transition line and the percolation transition line cross, which is similar to Ising spin systems where the percolation transition line and the order-disorder line meet on the critical point.
\end{abstract}


\label{sec:Introduction}
Colloidal particles are solid particles made from polymers, metals, and other various materials. Latex is an example of such colloidal particles. They show characteristic phenomena, for example depletion interaction and glass transition. It is known that such colloidal systems can be modelled by a rigid spherical particle system~\cite{Pusey}. The phase behavior of rigid particle systems shows a phase transition between the solid phase and the fluid one, which is called Alder transition~\cite{Alder1960,Alder1962}.

Since colloidal particles tend to aggregate and flocculate, polymers are often grafted on the colloidal particles used in industry to prevent such flocculation. They are called polymer-grafted colloidal particles. We can control the interaction between polymer-grafted particles by adjusting the length, species, and the grafting density of the grafted polymers and the quality of the solvent~\cite{Shull:1991,MatsenANDGardiner:2001,RoanANDKawakatsu:2002-1,BorukhovANDLeibler:2002}. Thanks to such usefulness, polymer-grafted colloidal particles can be adapted to a variety of applications and play an important role in industry. We study the effect of such grafted polymers on colloidal particles by Monte Carlo simulations using a simple model system. In the present article, we show that particles interacting via spherically symmetrical repulsive potential, without any attraction, self-assemble and finally align in strings, and that this string-like assembly possesses a similarity to the percolation transition and critical phenomena.

\label{sec:InteractionBetweenThePolymer-graftedColloidalParticles}
A pair of polymer-grafted colloidal particles interact not only with their rigid cores but also with their polymers outside the rigid cores. We can describe the interaction between the polymer-grafted colloidal particles by adding interaction potential of grafted polymers to hard core potential. For example, if the solvent is water, we graft block copolymers composed of a long hydrophobic block and a short hydrophilic block onto the colloidal particle onto which the end of the hydrophobic block is grafted. These grafted block copolymers form a melt brush covered with a thin swollen shell. Here a melt brush is defined as a brush without the solvent while a swollen brush is defined as a brush swollen by the solvent.

Since two swollen brushes prevent themselves from overlapping with each other due to the excluded volume effect, the interaction between a pair of swollen brushes is repulsive~\cite{FleerAndOthers:1993}. On the other hand, the interaction between two melt brushes is slightly attractive upon contact~\cite{FleerAndOthers:1993} since the chain conformation can relax when the brushes touch each other. The effect of the curvature of the grafting surfaces on the interaction between brushes can be evaluated using the so-called Derjaguin approximation~\cite{Israelachvili} as long as the radius of the curvature is small compared to the particle radius. However, if the radius of the curvature is the same order as the brush thickness, no good analytical result is known, and we have to rely on numerical self-consistent field (SCF) calculation~\cite{Shull:1991,MatsenANDGardiner:2001,RoanANDKawakatsu:2002-1,FleerAndOthers:1993,BorukhovANDLeibler:2002}.

\begin{figure}[!tbp]
  \centering
  \includegraphics[width=6.5cm,clip]{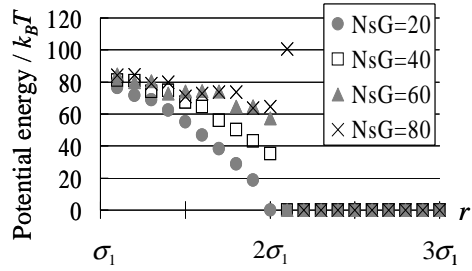}
  \caption{The results of the SCF calculations on the interaction potential between a pair of the polymer-grafted colloidal particles with diameter $\sigma_1$ for various polymerization index of grafted chains, $N_{SG}$. We set the size of a segment $0.1\sigma_1$, the polymerization index of the matrix chains $N_{SM} = 5.0$, the block ratio of the A-subchain of the grafted chains 0.998, the total volume of the segments of the grafted chains on both the colloidal particles $1.46\pi {\sigma_1}^3$, $\chi_{AB}=1.0$, $\chi_{AM}=1.0$, $\chi_{BM}=0.0$ where the indices A, B, and M indicate A segment, B segment, and the segment of the matrix chain respectively. At $N_{SG}=20$, these parameters correspond, for example, to the system obtained by grafting polypropylene with molecular weight $\approx 10^5$ onto colloidal particles $100\un{[nm]}$ in diameter with the graft density $\approx 1\un{[(nm)^{-2}]}$.}
  \label{fig:PotentialBySCF}
\end{figure}

First, we performed SCF calculations on two colloidal particles with diameter $\sigma_1$ onto which A-B block copolymers with polymerization index $N_{SG}$ are grafted~\cite{NorizoeForthcomingPaper:2005}, where A and B subchains are hydrophobic and hydrophilic respectively. Figure~\ref{fig:PotentialBySCF} shows the calculated interaction potential as a function of $r$, the distance between the centers of the pair of the particles. The result indicates that the interaction potential approaches the square-step potential when $N_{SG}$ increases while the total volume of segments of the grafted polymers is fixed.

We approximate this potential by spherically symmetrical step potential with a rigid core, i.e.
$\phi (r) = \infty \; (r < \sigma_1)$,
$\phi (r) = \epsilon _0 \; (\sigma_1 < r < \sigma _2)$,
and $\phi (r) = 0 \; (\sigma _2 < r)$,
where $\sigma_2$ is the diameter of the outer core formed by the polymer brush and $\epsilon _0 > 0$. Note that the height of the potential step $\epsilon _0$ is dependent on the temperature because it originates from the conformational entropy.


Owing to the step in the potential, the phase behavior of the system depends on temperature, which differs from the behavior of a rigid particle system~\cite{Alder1960,Alder1962}. The probability that a pair of particles get on the step in the potential is given by the Boltzmann factor $\exp (-\epsilon_0 /k_B T)$. The phase behavior at extreme temperatures is easily understood though such a temperature region is out of the validity range of our model potential.

At extremely high temperature the phase diagram of our system coincides with that of a system of rigid spherical particles with diameter $\sigma_1$ since the step in the interaction potential is neglegible compared to the kinetic energy.

At extremely low temperature, on the other hand, the step in the interaction potential becomes extremely high compared with the thermal energy. Owing to the potential energy barrier the phase diagram also coincides with that of a system of rigid spherical particles with diameter $\sigma_2$ when the system volume, $v$, exceeds the close-packed volume of the outer core, $v_0$. Then, the pressure jumps suddenly when the system is compressed across $v_0$, which leads to a phase transition.

In previous studies, a step potential with a rigid core (or one with both a rigid core and a square-well) was used as a model potential for cesium, cerium, water and so on to reproduce qualitatively the anomalies on their phase diagrams such as the anomalies in the behavior of the melting curves and the critical behavior of the phase diagram~\cite{DoubleStepPotential,Alder1979}. In two dimensions the model system forms dimers, lamellae, and other structures~\cite{Malescio:2004,Malescio:2003,Norizoe:2003April,Norizoe:2003November,Master'sThesis}.
In the present article we perform a series of simulations in order to obtain the complete phase diagram.

\label{sec:SimulationMethods}
In two dimensions Monte Carlo simulations based on the canonical ensemble have been carried out using the Metropolis algorithm~\cite{ComputerSimulationOfLiquids}. Mersenne Twister is adopted as a random number generator~\cite{MersenneTwister1}. In the initial state, $N$ particles are arranged on a triangular lattice in a box of volume (area) $v$ with periodic boundary conditions. At every simulation step a particle is picked at random and given a uniform random trial displacement within a radius of $0.2 \sigma_1$. A Monte Carlo step (MCS) is defined as every $N$ trial moves, during which each particle experiences one trial move on average. From now on dimensionless width of the potential step $\sigma_2 /\sigma_1$ is always fixed at 2.0. From beginning to end of one simulation, the number of particles $N$, the dimensionless volume (area) $v/v_0$ ($v_0 = N {\sigma_2}^2 \sqrt3 /2$ in 2 dimensions), and the dimensionless temperature $k_B T/\epsilon_0$ are fixed. After $5.0\times 10^6$ MCS, by which the system relaxes to the equilibrium state except at low temperature and high density, we acquire data every $10^4$ MCS and get 100 independent samples of particle configurations. In the present article, simulation results at $N=1200$ are mainly shown. We have confirmed, however, that the physical properties of the simulation system are essentially unchanged even when we simulate a larger system with $N=4800$.

\label{sec:SimulationResults}

\begin{figure}[!tb]
\centering
\includegraphics[width=8cm,clip]{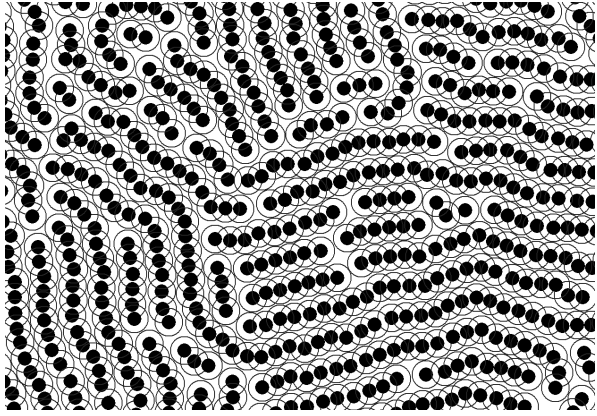}
\caption{The snapshot at $v/v_0=0.7$ and $k_B T/\epsilon_0=0.1$.}
\label{fig:first_result}
\end{figure}

As a particular example, we show a snapshot of the system at $N=9900$, $v/v_0=0.7$, $k_B T/\epsilon_0=0.1$ and $5.0\times 10^6$ MCS in Fig.~\ref{fig:first_result}. We observe that the colloidal particles start to self-assemble and finally align in strings. We call this effect ``frogspawn effect" on account of the characteristic configuration~\cite{Norizoe:2003April,Norizoe:2003November,Master'sThesis}. With use of the molecular dynamics simulation based on the micro-canonical ensemble at $N=140$, $v/v_0=0.7$, and internal energy per particle $=0.96 \epsilon_0$, which corresponds to $k_B T/\epsilon_0 \approx 0.036$, we checked that this string-like assembly is not an artifact of the Monte Carlo simulation and actually observed the same string-like assembly. This configuration in 2 dimensions is also observed in a model system with $\sigma_2 / \sigma_1 = 2.0$~\cite{Malescio:2004} or $2.5$~\cite{Malescio:2003}, where the system is cooled off from high temperature to low temperature, or in a model system that is composed of particles interacting via continuous repulsive potential~\cite{Camp:2005}.

\label{subsec:DefinitionOfStringAndCluster}
To discuss the structure of the aggregation, including string-like one, we define string length and cluster size. First we define a ``bond'' as an overlap between a pair of particles. When bonds form a single network composed of $M$ particles, we define this network as a ``cluster with cluster size $M$''. A particle that has more than two bonds in a cluster is defined as a ``junction''.

In Fig.~\ref{fig:first_result} we observe that clusters have different configurations of the network even if they have the same size. In order to quantitatively specify the network configurations of clusters, we introduce ``string length''. A string is defined as a sequence of particles, each of which has exactly two bonds. String length is defined as the number of particles along the string, including both ends, when the ends are junctions or particles with only one bond. We exclude a circular cluster composed of three particles from the definition of the strings though all the particles in such a cluster have two bonds.

\label{subsec:StabilityOfTheString-likeAssembly}
In order to check the dependence of the final particle configurations on the initial configurations, simulations with the same set of parameters have been performed from 4 different homogeneous and/or inhomogeneous initial configurations, which are perfect string-like assembly aligned in the $x$-direction, the close-packed configuration of the inner cores, etc.
%
We calculate the string length distributions and the cluster size distributions for these 4 cases. These two distributions characterize the strings and clusters. The calculated distributions show that all the simulations starting from different initial configurations end in the same string-like structure, whose cluster size shows exponential distribution. This result indicates that the string-like configuration is the stable configuration. The exponential distribution of the cluster size indicates that the clusters are results of independent aggregation events of particles. This is just like the formation of clusters in percolation system in its non-critical region.

In view of the internal energy we can understand the reason for the string-like assembly. At extremely low temperature the particles tend to get down the step in the potential to decrease internal energy. At $v/v_0 < 1.0$, however, the system does not have enough free volume to allow all the particles to get down the step. Consequently, at extremely low temperature and $v/v_0 < 1.0$, the particles form long strings since a string-like linear cluster takes lower internal energy than a concentrated bunching cluster does. As the temperature increases, the strings become short since the particles can climb the potential step more easily. Note that the particles do not form long strings at $v/v_0 \approx 1.0$ since the particles seldom overlap. At high density the particles compose only a few strings since the small free space allows very few particles to get down the step, and thus most particles are junctions.

\begin{figure}[!tb]
  \centering
    \includegraphics[height=5cm,clip]{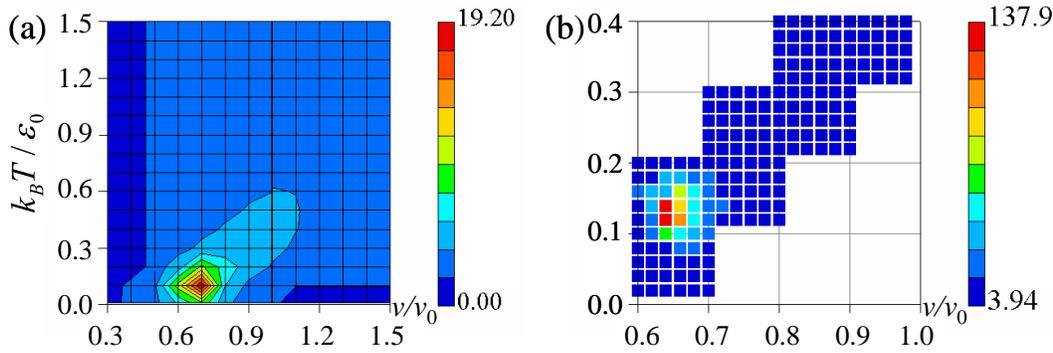}
  \caption{The average string length at $N=1200$. Blue regions represent short strings and red ones long strings.}
  \label{AverageSTlengthN1k}
\end{figure}

In order to determine the region of string-like configurations, we next calculate the number fraction of the particles that belong to any strings and the average string length. The average string length is shown in Figs.~\ref{AverageSTlengthN1k}(a) and (b). we observe that the average string length has a sharp peak at $0.1 < k_B T/\epsilon_0 < 0.2$ and $0.6 < v/v_0 < 0.7$. On the other hand, the calculated number fraction (data not shown in this article) shows wide regions of string-like assembly, located at $0.0 < k_B T/\epsilon_0 < 1.5$ and $0.3 < v/v_0 < 1.5$. These results indicate that there are many short strings in these wide regions outside this peak. At high density regions, however, there are only a small number of strings since very few particles can get down the step. At $v/v_0 > 1.0$ and low temperature regions particles can seldom overlap and most particles keep isolated. Figure~\ref{AverageSTlengthN1k}(b) also indicates that the maximum average string length exceeds 100 at this peak, which suggests a divergence of the string length in that region.

\begin{figure}[!tb]
  \centering
  \includegraphics[height=5cm,clip]{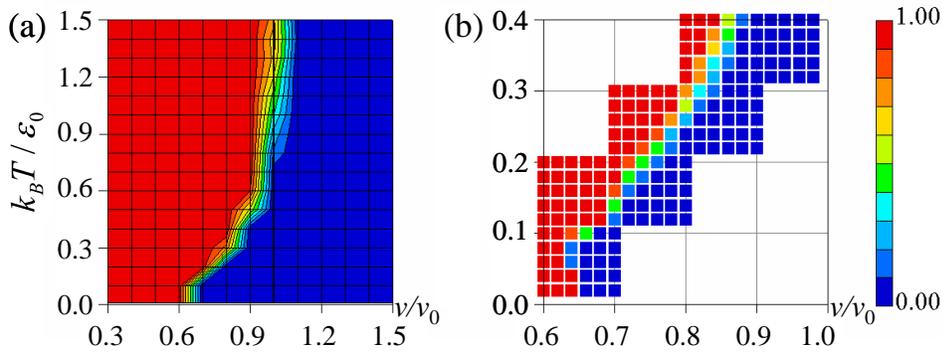}
  \caption{Occurrence probability of percolated clusters, (a), and its details around the boundary, (b), for the same systems as Fig.~\ref{AverageSTlengthN1k}.}
  \label{fig:PercolationN1k}
\end{figure}

At $N=1200$ and $k_B T/\epsilon_0=0.1$ cluster size distribution (data not shown in this article) suddenly changes between $v/v_0=0.6$ and $v/v_0=0.7$. At $v/v_0 = 0.7$ and $0.8$ the distribution has only one peak at cluster size 2, whereas at $v/v_0 = 0.6$ and $0.5$ a second peak appears around cluster size 1000. At $v/v_0 \geq 0.7$ a large majority of particles compose many small clusters in the system, whereas at $v/v_0 \leq 0.6$ most particles form one large cluster.
In order to understand the sudden change in the cluster size distribution, we evaluate the probability of the occurrence of a large cluster that bridges both sides of the system, i.e.~a percolated cluster~\cite{Stauffer1985}. In Fig.~\ref{fig:PercolationN1k}(a), we show the occurence probability of the percolated cluster that is obtained from the 100 independent samples at each point of the $(k_B T/\epsilon_0)$-$(v/v_0)$ diagram. We observe that the percolated clusters suddenly appear when the system crosses a sharp boundary around $v/v_0 \approx 0.7$-$1.0$. To determine the location of this boundary, we show in Fig.~\ref{fig:PercolationN1k}(b) a similar picture to Fig.~\ref{fig:PercolationN1k}(a), but with a finer resolution.
We can see that the occurrence probability of the percolated clusters changes abruptly on a sharp boundary, which suggests a similarity of this phenomenon to the usual percolation phenomena~\cite{Stauffer1985}.



The boundary in Fig.~\ref{fig:PercolationN1k}(a) asymptotically approaches to the line $v/v_0 = 1.0$ at high temperature, since at $v/v_0 > 1.0$ the system has enough free volume to allow all particles to get down the step and the particles do not frequently overlap. On the other hand, at low temperature percolation transition occurs at $v/v_0 \approx 0.65$, between Alder transition of outer cores and Alder transition of inner cores.

Now we study the relationship between our string-like assembly and the percolation phenomena. We calculate the cluster size distribution along the boundary curve in Figs.~\ref{fig:PercolationN1k}(a) and (b) (data not shown in this paper). The result shows that the cluster size distribution is a power law function of the cluster size with the exponent about 1.89. This exponent 1.89 corresponds to the value of the size distribution of percolated clusters on a 2-d triangular lattice, which equals 2.0~\cite{Stauffer1985}. Out of the percolation transition region, the cluster size distribution becomes an exponential function, as was already mentioned, which is also similar to the usual percolation phenomena.

Figures~\ref{AverageSTlengthN1k}(a) and (b) also show an anomaly, i.e. a divergence of the string length, around the boundary between the percolated region and the non-percolated region. This relation between string-like configuration and percolation transition indicates a similarity between our string-like assembly and the critical phenomena. Due to the poor resolution of the data, however, we cannot evaluate the critical exponent.

\begin{figure}[!tb]
  \centering
  \includegraphics[width=13cm,clip]{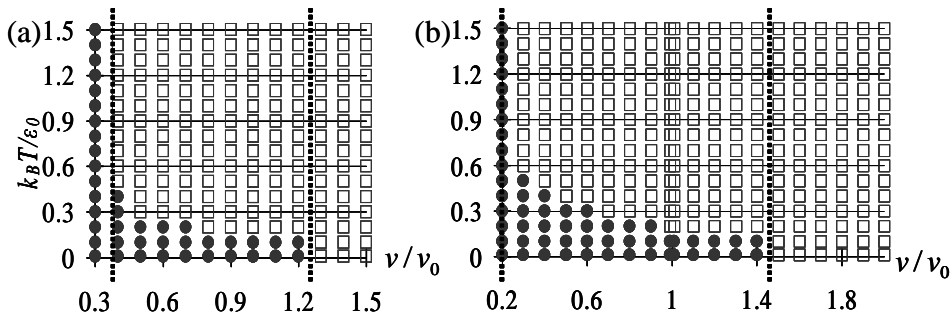}
  \caption{Phase diagrams in 2 dimensions at $N=1200$, (a), and in 3 dimensions at $N=4000$, (b). $\bullet$: Solid phase. $\square$: Fluid phase or coexistence of both phases. Dotted lines correspond to the Alder transition.}
  \label{fig:PhaseDiagram2Dand3D}
\end{figure}

By calculating the mean square displacement (MSD) of the particles during $10^6$ MCS, we determine regions of a solid phase and draw a phase diagram (see Fig.~\ref{fig:PhaseDiagram2Dand3D}(a)). We regard the system as a solid if the MSD value is less than its value at $k_B T/\epsilon_0 = 0.01$ and $v/v_0 = 1.3$. This is because Alder transition occurs at $v/v_0 \approx 1.3$ in 2D hard particle systems~\cite{Alder1960,Alder1962}.
\begin{figure}[!tb]
  \centering
  \includegraphics[height=5cm,clip]{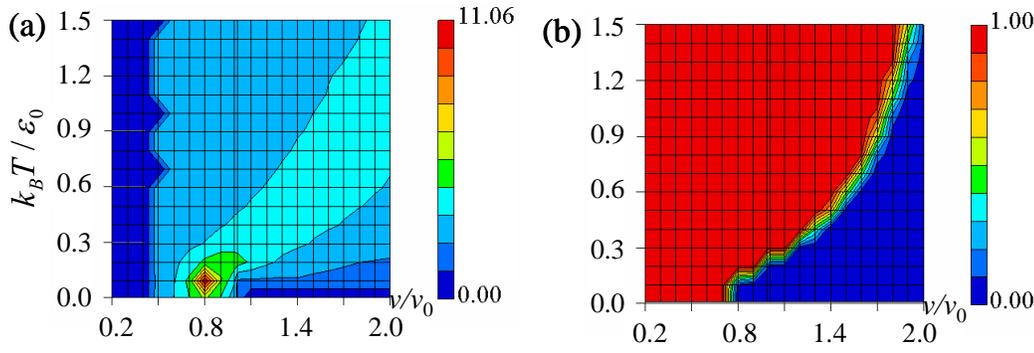}
  \caption{The average string length, (a), and occurrence probability of percolated clusters, (b), in 3 dimensions. $N=4000$. }
  \label{fig:Frog3DSimulationResults_EPL}
\end{figure}
At high temperature regions Alder transition of inner cores occurs at $0.3 < v/v_0 < 0.4$, whereas Alder transition of outer cores, which is expected at $v/v_0 \approx 1.3$, does not. At low temperature regions only Alder transition of outer cores emerges. At intermediate temperature regions the solid melts when not only the volume but also the temperature increases. These results correspond to the phase behavior discussed in the introduction.
The region where our string-like assembly like Fig.~\ref{fig:first_result} and percolation transition occur, i.e. $0.1 < k_B T/\epsilon_0 < 0.2$ and $0.6 < v/v_0 < 0.7$, is located in a solid phase between two Alder transitions. We can see that our string-like assembly like Fig.~\ref{fig:first_result} is an amorphous solid.

In three dimensions Monte Carlo simulations have also been performed according to the same simulation condition as 2D simulations, except that the particles are arranged on fcc lattice in the initial state and we acquire data after $10^6$~MCS. Figures~\ref{fig:PhaseDiagram2Dand3D}(b), \ref{fig:Frog3DSimulationResults_EPL}(a), and \ref{fig:Frog3DSimulationResults_EPL}(b) show a phase diagram, the average string length, and occurrence probability of percolated clusters in 3 dimensions respectively. At the percolation transition the cluster size distribution becomes a power law function of the cluster size with the exponent about 2.1, which is different from the value 1.89 for 2D case. These results qualitatively agree with the ones in 2 dimensions, which indicates that our discussion on 2D systems also applies to 3D systems. Our results in both 2 and 3 dimensions show that the average string length diverges around the region where the melting transition line and the percolation transition line cross. A similar behavior is found in Ising spin system where the percolation transition line and the order-disorder line meet at the critical point~\cite{Stauffer1985}.

In conclusion, using Monte Carlo simulations we studied phase behavior of 2 and 3 dimensional systems of hard particles with step repulsive interacting core. The simulation results show string-like assembly of particles between the two Alder transition densities, each of which corresponds to the inner hard core and the outer hard core. We found that this string-like assembly is related to the percolation phenomena. We can also observe the same string-like assembly like Fig.~\ref{fig:first_result} in 2 dimensional systems at $\sigma_2 / \sigma_1 = 1.9, 2.1$ and 2.5 though not at $\sigma_2 / \sigma_1 = 1.1, 1.5, 3.0, 3.5$ and 10.0, which will be discussed in our forthcoming paper~\cite{NorizoeForthcomingPaper:2005}.

\acknowledgments
The authors wish to thank Drs Nariya Uchida and Tsuyoshi Hondou, Mr Hiroaki Honda, and Prof. Toshiharu Irisawa for helpful suggestions and discussions. This work is supported by Grant-in-Aid for Science Nos.16340120 and 16654060 from the Ministry of Education, Culture, Sports, Science and Technology, Japan. The computation was in part performed at the Super Computer Center of the Institute of Solid State Physics, University of Tokyo.



\end{document}